\documentclass[aps,author-year,square,jor,twocolumn,a4paper]{revtex4-1}
\usepackage{ucs}
\usepackage[utf8x]{inputenc}
\usepackage{amsfonts}
\usepackage{amsmath}
\usepackage{amssymb}
\usepackage{amsbsy}
\usepackage[american]{babel}
\usepackage[T1]{fontenc}
\usepackage{subfigure}
\usepackage{verbatim}
\usepackage{graphicx}
\usepackage{xcolor}
\usepackage{url}
\usepackage{tensor}
\usepackage{lettrine}
\usepackage{hyperref}
\usepackage{color}
 \definecolor{darkgray}{rgb}{0.3,0.3,0.3}
 \definecolor{darkblue}{rgb}{0.0,0.0,0.4}
 \definecolor{darkred}{rgb}{0.4,0.0,0.0}
\hypersetup{
    unicode=true,          
    pdftoolbar=true,        
    pdfmenubar=true,        
    pdffitwindow=false,     
    pdfstartview={FitH},    
    pdftitle={Inertia-induced transient shear-banding in complex fluids},  
    pdfauthor={Marko Korhonen, Mikael Mohtaschemi, Antti Puisto, Xavier Illa, Mikko J. Alava},     
    pdfcreator={Marko Korhonen, Mikael Mohtaschemi, Antti Puisto, Xavier Illa, Mikko J. Alava},   
    pdfkeywords={wall slip, complex fluids, rheology}, 
    pdfnewwindow=false,      
    colorlinks=true,       
    linkcolor=darkred,          
    citecolor=darkblue,        
    filecolor=darkred,      
    urlcolor=darkgray           
}

\begin{document}

\author{Marko Korhonen}
\author{Mikael Mohtaschemi}
\author{Antti Puisto}
\author{Xavier Illa}
\author{Mikko J. Alava}
\affiliation{Aalto University, School of Science, Department of Applied Physics, P.O. Box 11100, FI-00076 AALTO, Finland}
\title{Start-up inertia as an origin for heterogeneous flow}
\begin{abstract}
For quite some time non-monotonic flow curve was thought to be a requirement for shear banded flows in complex fluids.
Thus, in simple yield stress fluids shear banding was considered to be absent.
Recent spatially resolved rheological experiments have found simple
yield stress fluids to exhibit shear banded flow profiles. One proposed mechanism for the
initiation of such transient shear banding process has been a small stress heterogeneity
rising from the experimental device geometry. Here, using Computational Fluid Dynamics methods, we show that
transient shear banding can be initialized even under homogeneous stress conditions by the fluid start-up inertia,
and that such mechanism indeed is present in realistic experimental conditions. 
\end{abstract}

\maketitle

\section{Introduction}
Soft Glassy Materials (SGMs) present a theoretical concept for fluids that can be solid-like or liquid-like at rest, but all
posses a microstructure with strongly interacting building blocks \cite{divoux2016}.
An extreme class of these are the yield stress fluids, which by virtue of their internal structure behave as solids under an imposed
mechanical stress below the yield point, yet flow like fluids once the stress is increased beyond this limit~\cite{moller2006yield,divoux2010transient,moller2009attempt}.
In addition to their extensive utilization in
commercial products as {\it{e.g.}} toothpastes, paints and process suspensions~\cite{coussot2002avalanche},
yield stress fluids are also of profound theoretical interest due to their rich and complex rheological behavior.
Indeed,  the peculiarities of flow response are also witnessed in {\it{shear banding}}, where the flowing fluid
exhibits a spatially banded structure, each band possessing a distinct viscosity associated with a unique shear
rate~\cite{manneville2008recent}.
This banding can appear as a true steady-state structure~\cite{olmsted2008perspectives} or as {\it{transient shear bands}}
(TSBs), which, while possibly extremely long lived, eventually are replaced by the homogeneous steady-state~\cite{moorcroft2011age}.
State-of-the-art observations find transient shear banding of simple yield stress fluids very robustly in different experimental
geometries \cite{divoux2010transient}.
Furthermore, even though non-thixotropic, such fluids display time- and shear rate or stress dependent response to step shear,
where the relaxation time apparently follows a power-law of both applied stress and shear rate \cite{Divoux2011}.

Steady-state shear-banding has been described as a property arising from the non-monotonic nature
of the fluid's intrinsic flow curve~\cite{olmsted2008perspectives,fielding2007complex}. This allows for a mechanically
unstable flow regime, where the negative slope in the constitutive curve permits the existence of a non-unique correspondence between
shear rates and shear stresses.
Thus, in this special regime the fluid may separate into two bands
of either identical shear stress (gradient banding) or identical shear rates (vorticity banding)~\cite{olmsted2008perspectives}.
Such a steady state scenario is unexpected for fluids known to have monotonic intrinsic flow curves~\cite{divoux2010transient}.

Against this reasoning, transient shear banding has recently been experimentally observed~\cite{divoux2010transient,C1SM05740E},
and theoretically predicted even in fluids possessing monotonic flow curves~\cite{adams2011transient}.
Numerous reasons for this type of banding have been proposed. 
In their studies, involving two models, the fluidity model\cite{fluidity-model} and spatially resolved version of the soft glassy rheology (SGR) model \cite{sgr-model}, 
Moorcroft and co-workers~\cite{moorcroft2011age} predicted transient shear banding resulting from a mechanical instability
in the start-up flows. This was clearly seen in the instantaneous (stress-time) constitutive curve as a stress overshoot and
subsequent negative slope of the curve as it regresses towards the steady state.
In a more recent work, the same authors derive a more rigorous expression for the onset of this mechanical instability in multiple
flow scenarios~\cite{moorcroft2014shear}.

Transient shear banding in amorphous solids has been successfully modeled by Shi and Falk with
shear transformation zone (STZ) theory, in which the plastic deformations occur in specific zones, activated by an effective
temperature~\cite{shi2005strain}. At present, Hinkle and Falk are extending this framework to study
transient shear banding in simple Yield Stress Fluids (YSF)~\cite{hinkle2015rate}. These studies properly explain the movement and limited life-time
of the shear bands, however, they leave open detailed origin of the band initiation. To study this aspect,
a slightly different approach was proposed in~\cite{illa2013transient},
where a phenomenological, simple scalar $\phi$-model is utilized. In this case, the local microstructure of a time-dependent fluid is
encoded in the $\phi$ parameter, which refers to the local immobilized volume fraction of the fluid. In the framework
of this model, transient shear banding can initiate due to the small stress gradients that are induced by the rotational Couette flow geometry,
often utilized in experimental studies. These stress variations
generate shear rate heterogeneities, which in turn result in the non-uniform disintegration of the underlying fluid structure,
observed as the formation of a TSB~\cite{illa2013transient,lehtinen2013transient}. However, in experimental geometries,
such as the cone-and-plate, where TSB is also found, the stress heterogeneity is considered to be negligible. 

In this work, we consider the role of fluid inertia to initiate transient shear bands during the start-up, by solving
the Navier-Stokes equations for incompressible flow simultaneously to the evolution of a scalar field variable describing
the complex fluid structure. The purpose is to show the effect of accelerating flow to the transient shear band formation,
neglected earlier.
As reviewed above, to generate transient shear bands, some models pre-initialize by hand a non-homogeneous structure
profile~\cite{hinkle2015rate,adams2011transient} and others rely on the geometrical stress heterogeneity~\cite{illa2013transient,adams2011transient}.
While especially the latter mechanism compares reasonably to most experimental setups
(Couette, parallel plates), its relevance in the shear band initialization may be questioned in others (cone-and-plate). 
Our results suggest a plausible mechanism initializing transient shear bands even under perfectly homogeneous
initial conditions due to the flow acceleration mechanism.

Utilizing a Computational Fluid Dynamics (CFD) approach, we apply the $\phi$-model as in Ref.~\cite{illa2013transient}
in a homogeneous flow (planar Couette) scenario, where, as mentioned, any and all stress heterogeneities are due to inertial
effects alone.
We emphasize here, that, as will be explained later, the key concept in this model is the fact that the structure breakdown
is proportional to the shear rate, a fact based on advanced experiments~\cite{divoux2010transient}.
Thus, the qualitative behavior shown here, is not influenced by this particular model, but is general for all the
models and real physical materials that have this property.
The paper is organized as follows: first, we detail the theoretical framework and the implementation. Then,
we proceed to comment on the relevant results obtained with our approach and finally, the paper finishes with concluding
remarks.

\section{The model}
As the aim here is to examine TSBs in a time-dependent fluid, a necessary 
prerequisite for modeling transient shear-banding theoretically involves coupling a
structural model describing the internal structure of the fluid to the subsequent flow dynamics. Additionally, as the focus
is on providing a minimal example that presents qualitative evidence of transient shear-banding
driven by inertial effects, this structural model is chosen from a class of simple phenomenological models
based on the abstract scalar
structural parameter~\cite{coussot2002viscosity,coussot2002avalanche,cheng1965phenomenological,cheng2003characterisation,mujumdar2002transient,illa2013transient}.
The specific $\phi$-model addresses the internal structure of the fluid through the 
structural parameter $\phi$, describing the immobilized volume fraction present in the fluid~\cite{illa2013transient}.
This model allows fine-tuning the fluidization exponent to match the experimental values of various complex fluids, while other similar scalar models (such as the $\lambda$-model by Coussot {\it{et al}}.~\cite{lehtinen2013transient}) inherently produce a fixed fluidization exponent, therefore
limiting their applicability for this purpose. Indeed, to the best of our knowledge, the $\phi$-model is the simplest one possessing this crucial feature. Furthermore, the fluidization behavior of more elaborate models, such as the SGR-model~\cite{moorcroft2013criteria,sollich1997rheology}, are presently not reported.
In the $\phi$-model, the progression of the immobilized volume fraction, $\phi$, in time can be due to shear (constructive and destructive) and
shear-independent motion of the structural elements (constructive), reflected in the model as two superimposed
kernels. The time-evolution equation for $\phi$ is~\cite{illa2013transient}
\begin{equation} \label{eq:PhiModel}
 \frac{d\phi}{dt} = \frac{A_b}{(\mu/\mu_0)^m} + (A_s - B_s\phi)\left(\frac{\dot{\gamma}}{\dot{\gamma}_0}\right)^k ,
\end{equation}
where $A_s$ ($B_s$) is a kinetic constant for the shear growth (destruction),
$\dot{\gamma}$ is the magnitude of the strain rate (shear rate) and both $k$
and $\dot{\gamma}_0$ describe the sensitivity of $\phi$ towards shearing. Furthermore,
$A_b$, $\mu_0$ and $m$ describe the growth of $\phi$ due to shear-independent effects.
Again pursuing a minimal example, setting $A_b = 0$ in Eq.~\eqref{eq:PhiModel} and defining the sample history by fixing the
initial immobilized volume fraction $\phi_0$ allows for a minimal inspection of an internally
structured fluid with a steady-state $\phi$ of $\phi_{ss} = A_s / B_s$, with the structure breakdown rate
set by $B_s$ and proportional to the powers of the dimensionless shear rate $(\dot\gamma/\dot\gamma_0)^k$.
The proportionality $A_s/B_s$ gives the minimum volume fraction reached by a
physical system. For instance in an aggregating colloidal suspension, this can be thought to
describe the monomer volume fraction or in a microgel this could be the volume fraction at the maximally compressed state of the sponge like elements. We define $0 \leq A_s \leq B_s \leq 1$. The parameter  $\dot\gamma_0$ determines the system sensitivity to shear; a larger number means longer relaxation time.

The $\phi$-parameter can be incorporated into the flow quantities, as done here, by the
empirical Krieger-Dougherty relation~\cite{illa2013transient,lehtinen2013transient}
\begin{equation} \label{eq:KD}
 \eta(\phi) = \eta_0 \left(1-\frac{\phi}{\phi_m}\right)^{-n} ,
\end{equation}
where $\eta$ is the viscosity of the fluid, $\eta_0$ describes the viscosity of the suspending
matrix and $\phi_m$ denotes the jamming volume fraction, which describes a completely jammed
configuration.  This gives an additional constraint to the kinetic constants so that $A_s/B_s < \phi_m$.
The quantities used in Eq.~\eqref{eq:KD} are set to $\phi_m = 0.68$,
$\eta_0 = 1\ \mathrm{mPas}$ (water) and $n = 1.82$~\cite{illa2013transient}.

The final step in the model development is to include Eq.~\eqref{eq:KD} in the Navier-Stokes
equations that describe the flow field completely. The incompressible Navier-Stokes equations read
\begin{equation} \label{eq:NavStoMass}
 \nabla \cdot \mathbf{v} = 0 ,
\end{equation}
and
\begin{equation} \label{eq:NavSto}
 \frac{\partial \mathbf{v}}{\partial t} + \mathbf{v} \cdot \nabla\mathbf{v} = -\frac{\nabla p}{\rho} + \nu\nabla^2\mathbf{v} + \frac{\mathbf{f}}{\rho} ,
\end{equation}
where Eq.~\eqref{eq:NavStoMass} is incorporated in the latter equation, physically implying the conservation of momentum. In these equations,
$\mathbf{v}$ is the velocity field, $p$ is the pressure, $\rho$ is the density of the fluid and $\mathbf{f}$ describe bodily forces
({\it{e.g.}} gravity) acting on the fluid (here, $\mathbf{f}=0$). Additionally, $\nu$ is the kinematic viscosity of the fluid, readily obtained
from Eq.~\eqref{eq:KD} applying the relation $\nu = \eta / \rho$. The inertial effects,
generally omitted in rheological modeling, which usually deals with creeping flow conditions,
are included in the convection and acceleration terms on the left-hand side of Eq.~\eqref{eq:NavSto}.
\begin{figure}[h]
 \includegraphics[width=0.48\textwidth]{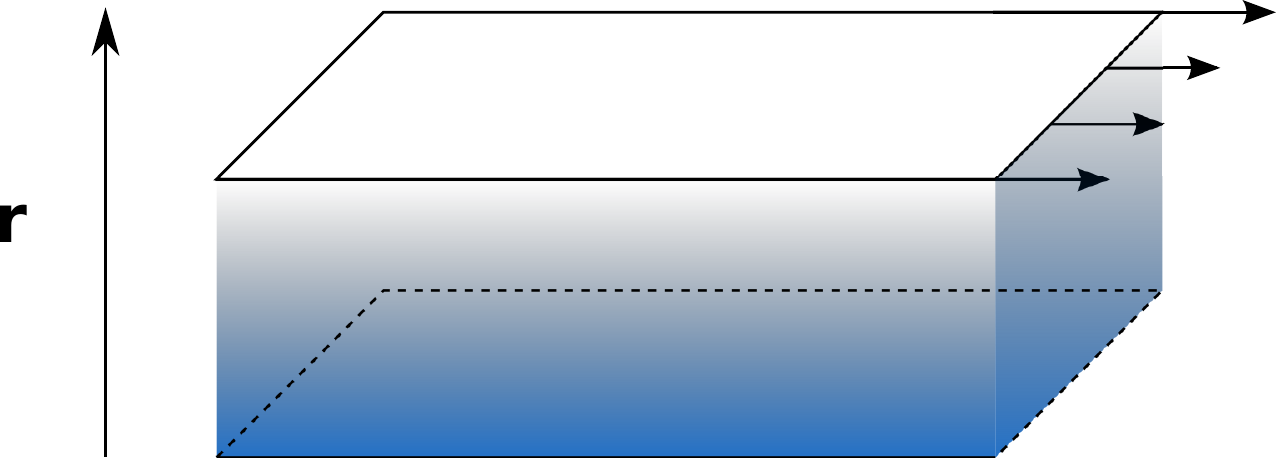}
 \caption{The planar Couette flow geometry used in this work. The upper plate induces simple shear to the fluid
 confined between the two plates, while the lower one remains stationary. Ideally, the stress is uniform over the
 gap between the plates.}
 \label{fig:Geometry}
\end{figure}

Transient shear-banding emerges as a
property in the fluidization of an internally structured fluid experiencing spatial stress variations and subsequent shear rate
inhomogeneities, leading to spatially non-uniform structural disintegration. As the viscosity is assumed a function
of the structural integrity (here modeled by coupling Eqs.~\eqref{eq:KD} and~\eqref{eq:PhiModel}), this is experimentally
seen as viscosity bands across the measurement gap. Accordingly, to isolate any sources other than the inertial contributions
for such variations, the planar Couette geometry shown in Fig.~\ref{fig:Geometry} is utilized in this work.
The fluid, confined between the two parallel plates, is subjected to simple shear as the upper plate moves and the lower one
remains stationary. In this scenario, the stress is uniform over the gap between the plates, ensuring that all stress variations
are due to inertial effects 
alone.

This approach is implemented in a Computational Fluid Dynamics (CFD) environment.
The flexible OpenFOAM\textregistered~software package~\cite{weller1998tensorial,chen2014openfoam} that
includes optimized numerical solvers, meshing, and visualization tools is applied and extended to account for the scenario presented here.
The simple shear geometry of Fig.~\ref{fig:Geometry} is meshed to a 2-D 60$\times$60 grid with periodic boundary conditions at the open
boundaries. The practical implementation extends the capabilities of nonNewtonianIcoFoam solver to allow for time-dependent viscosity.
The system of differential equations, is solved using the PCG (PCBiCG) linear solver with DIC (DILU) preconditioning and smoothing,
setting the absolute tolerance close to the machine precision of the 64-bit system ($10^{-16}$).
The kernel constants of the $\phi$-model in Eq.~\ref{eq:PhiModel} were set to $A_b=0$, $A_s=0.56$, $B_s=1.0$ and $\dot{\gamma}_0=1$.
Additionally, the gap width $e$ was set to $e=1.0$ mm. For brevity, the following results refer to the normalized
gap (scaled by $1/e$) when necessary.
The qualitative results are independent of the selection of these parameters. However, varying the $\dot\gamma_0$ shifts the fluidization curves (see Fig. 4) in the horizontal direction. Changing the ratio $A_s/B_s$ to a smaller value improves the contrast of the shear band edge, i.e. makes the kink in the velocity profile steeper, and shifts the fluidization curves in the vertical direction having similar effect as increasing the $\phi_0$ (Fig. 4).

\section{Results}
\begin{figure}[h]
 \includegraphics[width=0.49\textwidth]{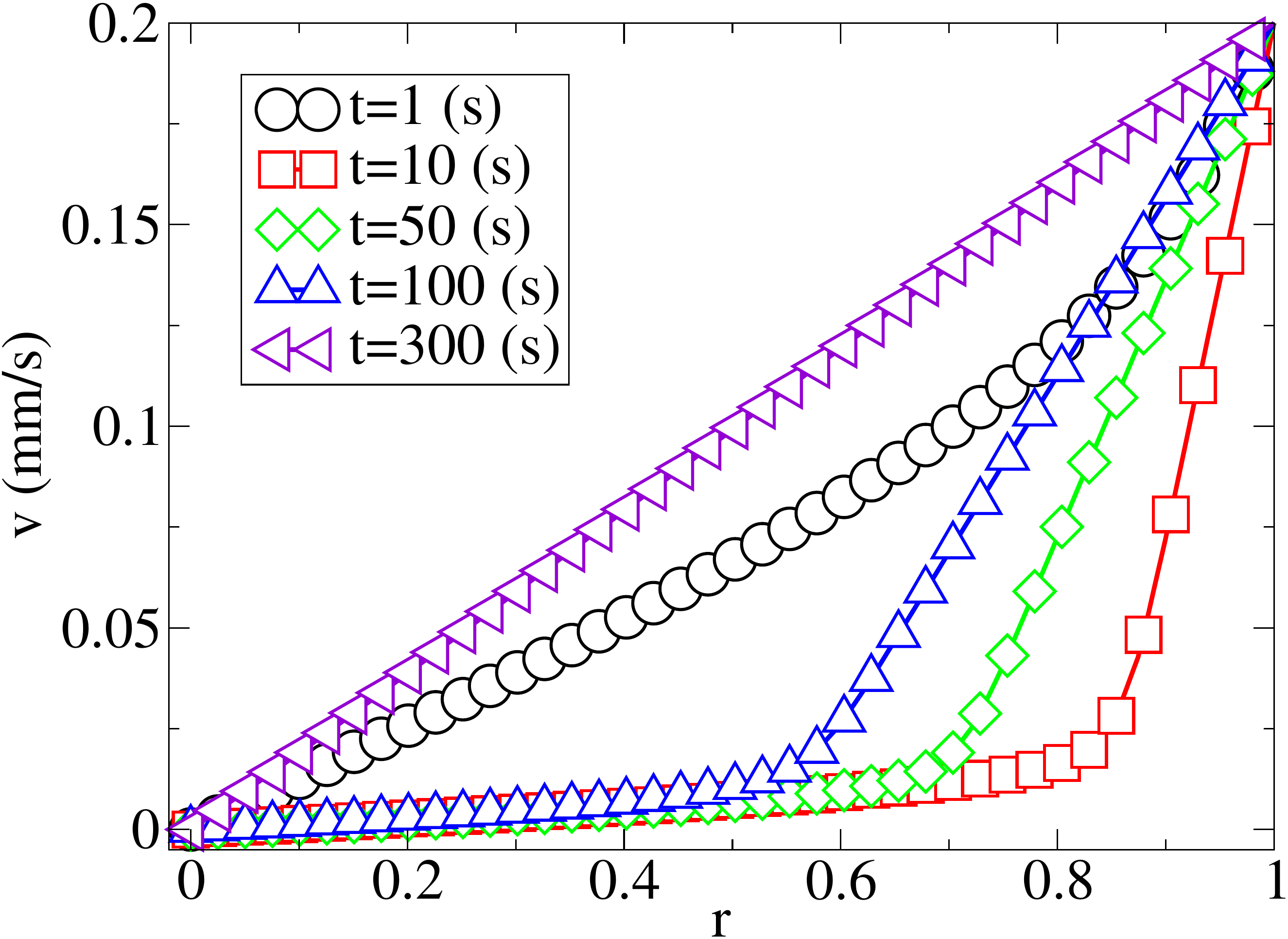}
 \caption{Planar Couette start-up flow: linear velocity profile.
 The model parameters are $A_s=0.56$, $B_s=1.0$, $k=1.6$ and $\phi_0=0.67999$ and the shear rate is fixed at $\dot{\gamma}=0.2\ \mathrm{1/s}$.
 The stress variations induced by the inertial effects result in the formation of a TSB, seen here as the initial shear localization
 towards the shearing (outer) plane. Gradually, the profile evolves over time towards the steady-state solution ($t=300$s) and the TSB
 vanishes.}
 \label{fig:fig1}
\end{figure}

\begin{figure}[h]
 \includegraphics[width=0.49\textwidth]{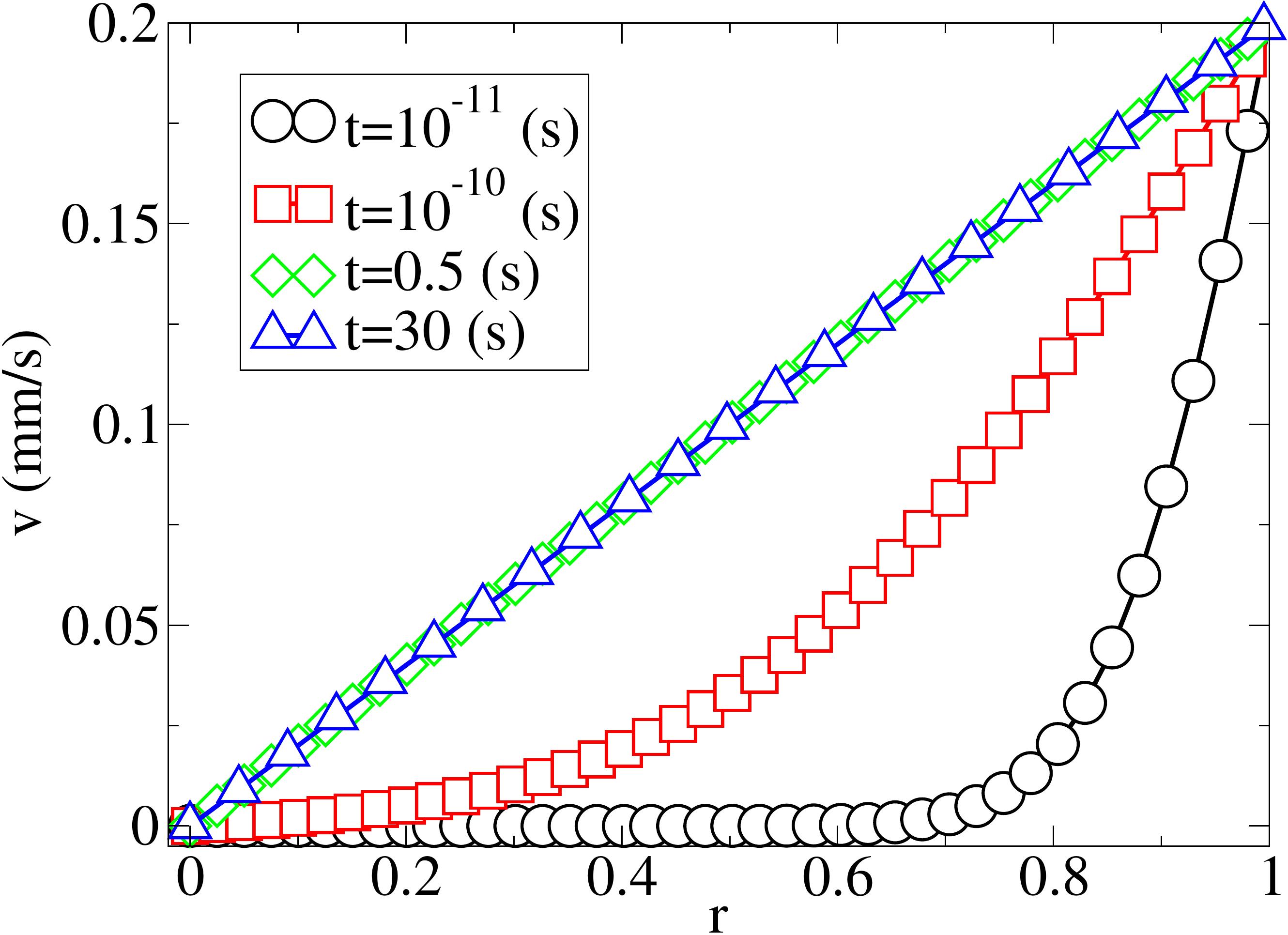}
 \caption{Planar Couette start-up flow: linear velocity profile.
 The model parameters are $A_s=0.56$, $B_s=1.0$, $\phi_0=0.67999$ and the shear rate is fixed at $\dot{\gamma}=0.2\ \mathrm{1/s}$ while the exponent of the $\phi$-model
 is now reduced to $k=1.0$. Here, the stress variations induced by the inertial effects are unable to produce a TSB due to the highly linear
 breakdown of the underlying fluid structure.}
 \label{fig:fig3}
\end{figure}

 The Navier-Stokes equation is characterized (at steady state) by the Reynolds number, which is represented here by the shear rate. In addition in the relaxation of the model depends on the shear rate exponent $k$ and the initial condition $\phi_0$. Thus, this set of parameters define the TSB and is studied in what follows.
In Fig.~\ref{fig:fig1} ($\phi_0=0.67999$, $\dot{\gamma}=0.2\ \mathrm{1/s}$), the linear velocity profile
is displayed in the homogeneous, planar Couette start-up flow scenario. In experiments, these 1-D pictures
are accessed by {\it{e.g.}} ultrasonic speckle velocimetry~\cite{divoux2010transient} and, in general, are used to identify shear
localization effects. 
Observing the evolution of the velocity in the flow direction allows to monitor the shear localization here, as well.  
As demonstrated in the example scenario of Fig.~\ref{fig:fig1}, shear banded flow profile develops in the course of a few seconds from
the on-set of the flow. Here we emphasize that no external initial disturbance to the model is required  and the shear localization is induced by the stress inhomogeneity due to the initial acceleration of the fluid, which
interacts with the viscosity evolution as discussed in detail for the creeping flow case in Ref. \cite{illa2013transient}.
With these model parameters, the TSB persist for up until $t\approx300$ s where complete fluidization and the steady-state
flow profile is reached. This certainly is a time-scale comparable to the experiments, where long lasting transient
banding was observed~\cite{divoux2010transient}. 
Note that since the mechanism here is based on the flow model interaction with the time-dependent shear thinning of the fluid, any
model having this property will potentially experience shear banding induced by the same effect. Experimentally this behavior is
present in systems showing microstructural disintegration due to shear, where the time-dependence is a power-law (typically of exponents
in the range of 2-4) of the shear rate or shear stress~\cite{divoux2010transient}.
Conversely, if the structural decomposition is linear (power-law exponent $k$ $\approx$ 1.0),
the inertial effects and the subsequent stress
variations are incapable of generating a TSB. This is illustrated in Fig.~\ref{fig:fig3}, where the model parameters are identical to those in Fig.~\ref{fig:fig1}
aside from the exponent of the $\phi$-model ($k=1.0$).
Here, due to the linear dependence between $\phi$ and $\dot{\gamma}$, the structural breakdown is uniform and
the shear localization does not persist up to the time scales observed in Fig.~\ref{fig:fig1}. Instead, the fluidization time is simply established
by the viscous time scale $\tau_v=\rho w^2 / \eta$ ($w$ denoting the gap width), as for a simple (Newtonian) liquid. Judging by this expression,
the high viscosity of the jammed initial configuration should yield an extremely low fluidization time. Indeed, as seen in Fig.~\ref{fig:fig3}, the initial
fluidization occurs within $10^{-10}$ seconds from the moment the flow commences.

\begin{figure}[h]
 \includegraphics[width=0.49\textwidth]{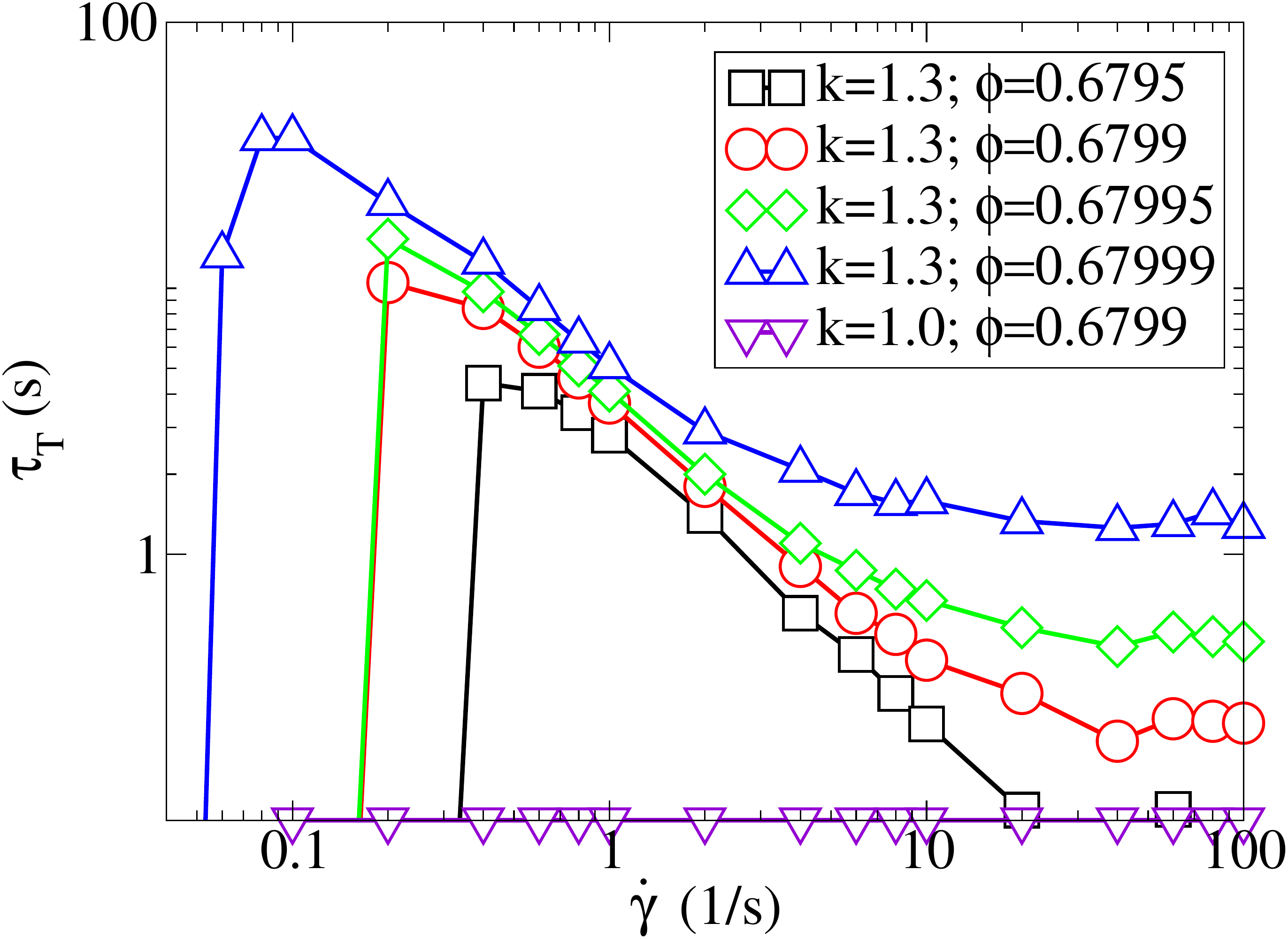} 
 \caption{Planar Couette start-up flow: fluidization time $\tau_f$ vs. applied shear rate $\langle \dot{\gamma} \rangle$ on a log-log scale.
 A sufficient applied shear rate is required for shear localization and nonuniform fluidization, seen as a discrete ramp in the results.
 The fluidization time decays rapidly by increasing $\langle \dot{\gamma} \rangle$.}
 \label{fig:fig2}
\end{figure}

On the other hand, in the model, shear localization occurs only on certain conditions:
the formation of a TSB is largely dependent on the model parameters as in our earlier work~\cite{illa2013transient},
where both the values of sample history variable $\phi_0$ and the kinetic exponent $k$ are able to predict the transient shear banding.
However, here, the imposed shear rate $\langle \dot{\gamma} \rangle$ also influences this process,
since it is this quantity that provides the necessary stress heterogeneity in the beginning of the flow.
This point is illustrated in Fig.~\ref{fig:fig2}, where the fluidization time $\tau_f$ (here, the time the TSB persists) is represented as a function
of the applied shear rate $\langle \dot{\gamma} \rangle$. For each $\langle \dot{\gamma}\rangle$,
$\tau_f$ was determined from the linear velocity profile (as shown in Fig.~\ref{fig:fig1}) as follows:
two separate linear curves were fitted to the profile, one spanning the area $0 \leq r/e \leq 0.33$
and the other $0.66 \leq r/e \leq 1.0$.
Once the slopes ($a_1$ and $a_2$) of these curves
reached a certain threshold (here: $|a_1-a_2|/a_2 \leq 5.0\cdot10^{-2}$), the profile was determined to having reached
the homogeneous flow state. The results are displayed in Fig.~\ref{fig:fig2}, where it is observed that
a sufficient value of $\langle \dot{\gamma}\rangle$ (and accordingly, sufficient stress heterogeneity)
is required for transient shear banding, otherwise an uniform breakdown of the underlying structure is witnessed
($\tau_f=0$). Once this $\langle \dot{\gamma}\rangle$ is reached, the time required for complete fluidization decays rapidly.
Moreover, if the structural decomposition is linear ($k=1.0$), the breakdown is uniform regardless of $\langle \dot{\gamma}\rangle$.
Note, that the discrete ramp in the fluidization times should not be interpreted literally as
the transformation from strong shear localization (TSB) to homogeneous
flow is of continuous nature. However, to quantify the fluidization times, we have resorted to the procedure introduced above,
which essentially imposes a cut-off criterion (and the discrete jump) to the results.

\begin{figure}[h]
 \includegraphics[width=0.49\textwidth]{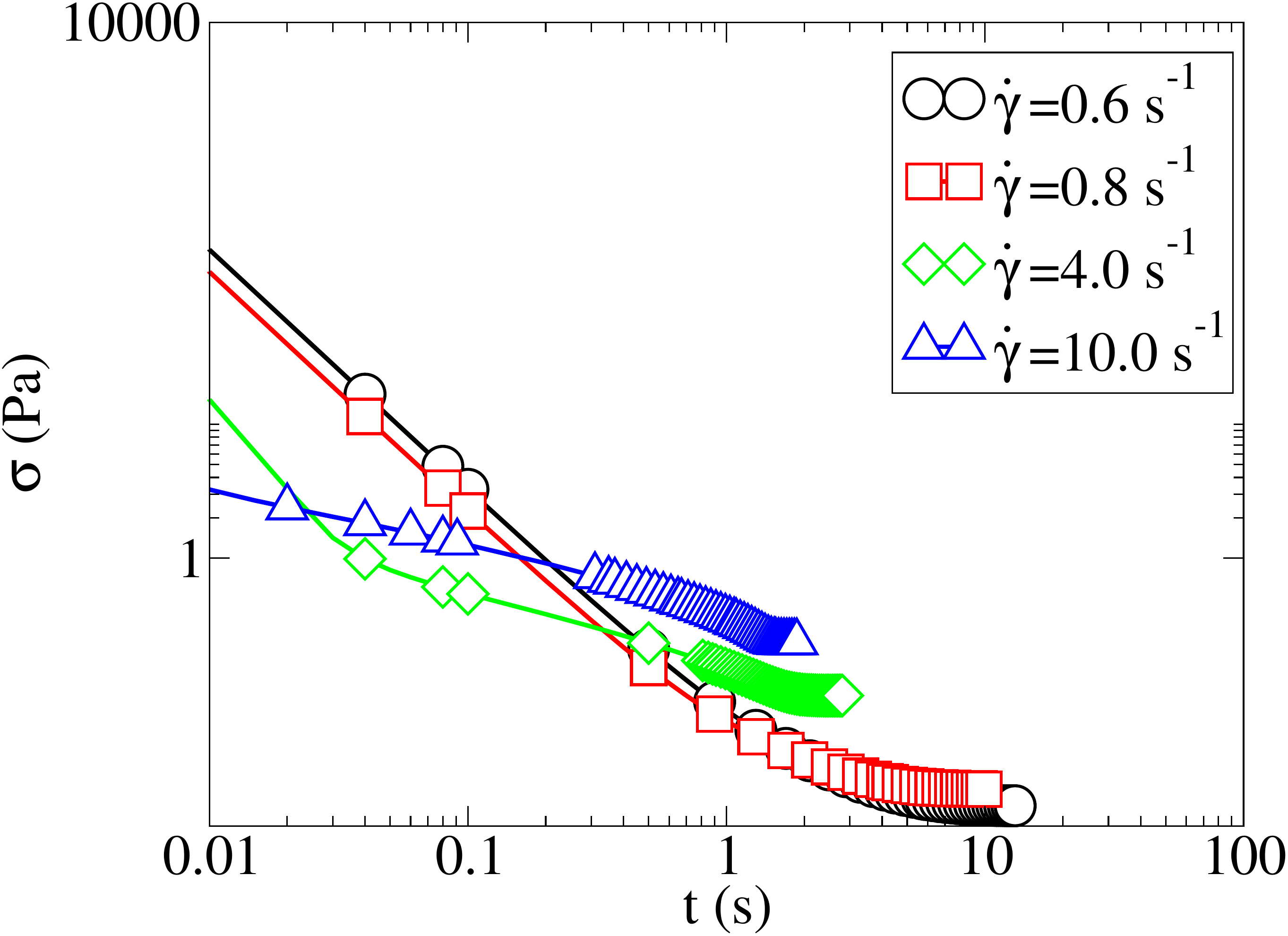} 
 \caption{Planar Couette start-up flow: the instantaneous stress-time curve ($A_s=0.56$, $B_s=1.0$, $k=1.3$ and $\phi_0=0.6799$).
 As time progresses, the structure breaks down, and a lower stress is required to maintain the applied shear rate 
 $\langle \dot{\gamma} \rangle$. If the magnitude of $\langle \dot{\gamma} \rangle$ is sufficient for forming a TSB,
 the corresponding stress-time curve exhibits a distinct kink (as seen here for $\langle \dot{\gamma} \rangle$ =  and
 $\langle \dot{\gamma} \rangle$), signaling the beginning of the shear localization. In the event of weak or nonexistent
 localization, this instability is not observed.}
 \label{fig:fig4}
\end{figure}

\begin{figure}[h]
 \includegraphics[width=0.49\textwidth]{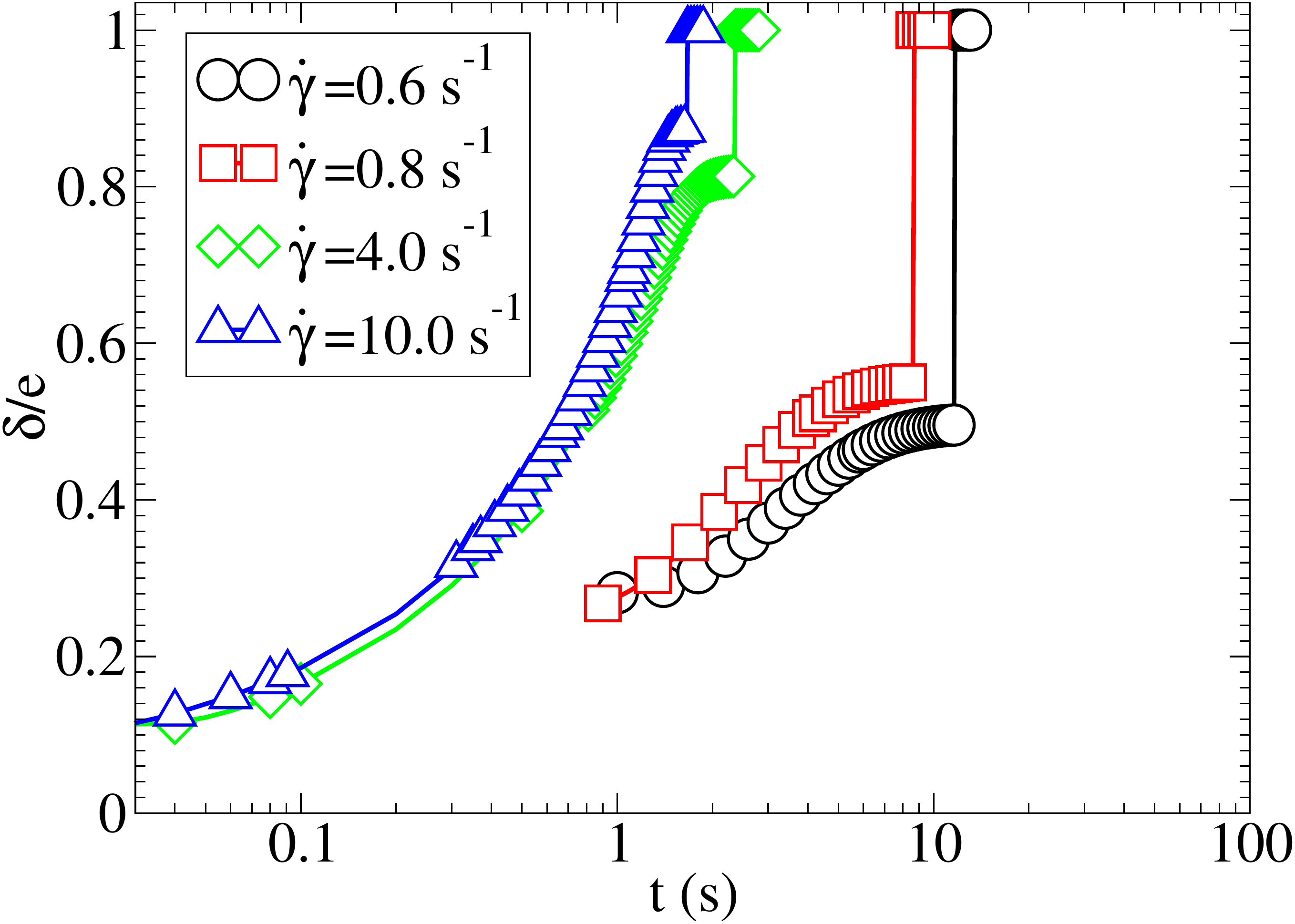} 
 \caption{Planar Couette start-up flow: the shear-band edge location (normalized gap) over time ($A_s=0.56$, $B_s=1.0$, $k=1.3$ and $\phi_0=0.67999$),
 calcucated for various applied shear rates.}
 \label{fig:fig5}
\end{figure}

Additionally, a complementing point of view on TSB formation is provided by various temporal plots. In Fig.~\ref{fig:fig4}, the apparent shear
stress $\sigma_a$, measured at the shearing plane, is displayed as a function of time $t$ for various applied shear rates $\langle \dot{\gamma} \rangle$. As the flow
progresses with increasing time, the microstructure of the fluid disintegrates and the local jammed volume fraction at the shearing plane decreases, leading
to a monotonic decrease of the effective viscosity as well, as the two are linked by Eq.~\eqref{eq:KD}. This, in turn, leads to a monotonic decrease
in the shear stress, as observed in the figure. Furthermore, the time at which the stresses reach their steady-state values
seems to concur well with the fluidization time observed in Fig.~\ref{fig:fig2} as one would expect.
Another temporal plot is illustrated in Fig.~\ref{fig:fig5}, where the shear band edge is tracked over the normalized gap ($\delta/e$) as a function
of time $t$. In this plot, the value $\delta/e=0$ corresponds to a scenario where the flow is completely localized in the region near the shearing
plane and $\delta/e=1$ where the TSB vanishes. For consistency, the shear band edge $\delta$ was determined by the linear fitting scheme
introduced earlier (see the commentary on Fig.~\ref{fig:fig2}). The edge location was determined as the intersection point of the two linear curves
as long as the steady-state condition ($|a_1-a_2|/a_2 \leq 10^{-2}$)) was not met. In the framework of the $\phi$-model, the shear band edge
travels at a varying (non-constant) velocity, dependent on the applied shear rate $\langle \dot{\gamma} \rangle$. In addition, the fluidization process is gradual up to a critical point, at which
the fluid suddenly yields completely. This is observed in Fig.~\ref{fig:fig5} as the discrete jump of the edge location to the value $\delta/e=1$.
The exact location of this jump is dependent on the applied shear $\langle \dot{\gamma}\rangle$. The behavior witnessed here should be compared with the corresponding earlier results~\cite{illa2013transient},
where the shear band edge also proceeds in a highly non-linear manner. However, the critical point is not observed, and the edge travels independent of
the applied shear rate $\langle \dot{\gamma} \rangle$ for all $k$. The reason for these differences lies in the fundamentally different source of the
stress gradient, that serves as the catalyst for transient shear banding. In Ref.~\cite{illa2013transient}, this stress variation is induced by the geometry and is essentially time-independent.
Here, the stress variation is a (time-dependent) result of the finite fluid inertia, which is directly influenced
by $\langle \dot{\gamma} \rangle$. This leads to a weaker localization of the fluid flow at low $\langle \dot{\gamma} \rangle$ and strong localization at high $\langle \dot{\gamma} \rangle$. Indeed, the blue
curve in Fig.~\ref{fig:fig5} ($\langle \dot{\gamma} \rangle = 10$s$^{-1}$) bears a strong resemblance to the results in Ref.~\cite{illa2013transient}, while lower values of $\langle \dot{\gamma} \rangle$
yield a more convoluted profile: the velocity of the shear band edge increases at first, but then decreases and finally diverges as the
critical point is reached. Thus, rather than reaching the static plate, the shear band disperses suddenly as the stress gradient has diminished over time and is unable to sustain
the banded profile.

\begin{figure}[h]
 \includegraphics[width=0.49\textwidth]{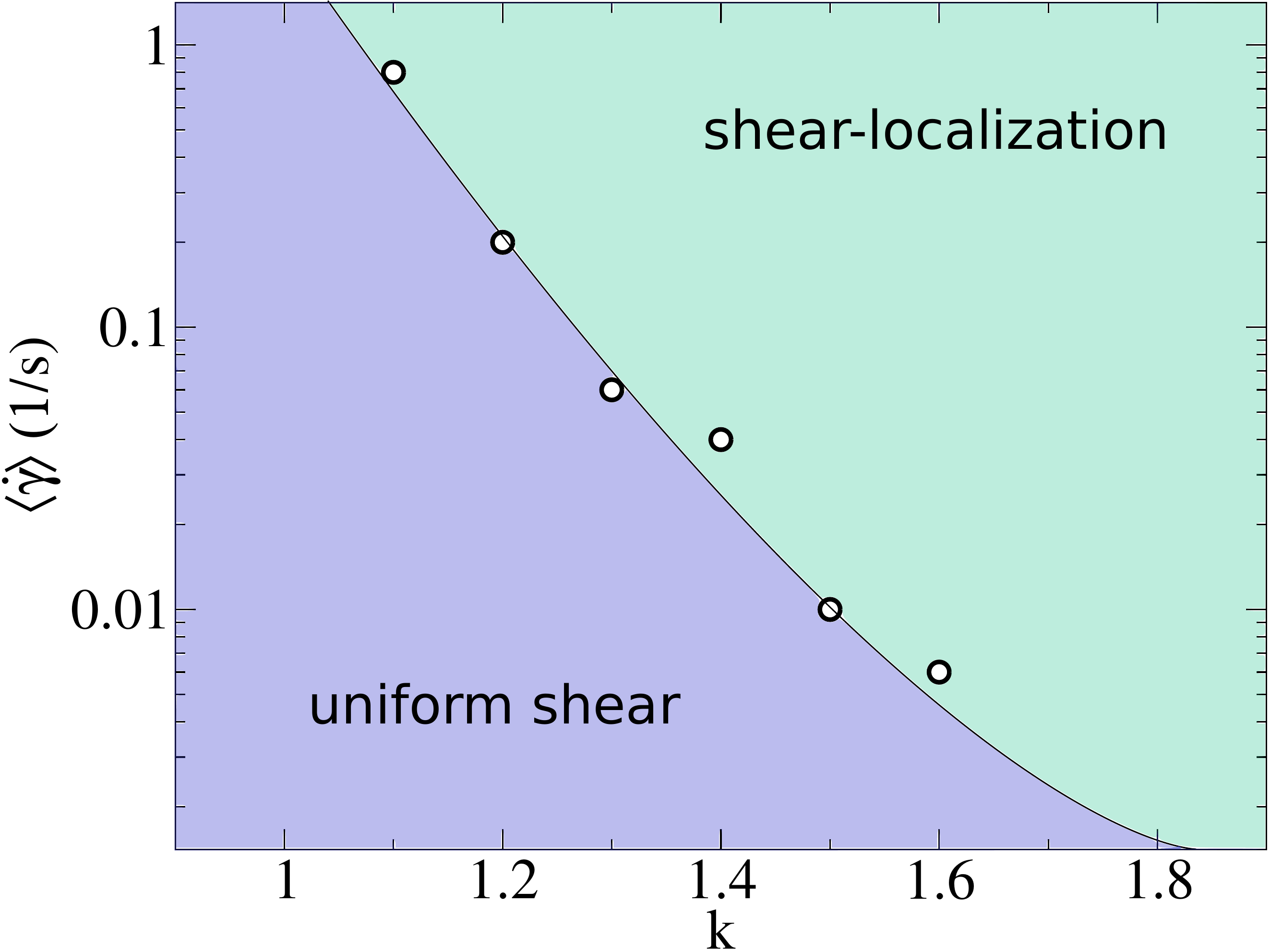} 
 \caption{Planar Couette start-up flow: a semilogarithmic phase diagram which classifies the flow as either uniform or localized (a TSB is formed).
 The boundary is defined by the minimum (triggering) value of the applied shear rate $\langle \dot{\gamma} \rangle$ (dependent on $k$), above which
 the start-up inertia induces a TSB and the corresponding banded profile is retrieved. Below this value, the flow remains decidedly uniform.
 As seen here, this value declines rapidly as a function of $k$. Note that the $k$ values for real-world complex fluids are on the order of $k \sim 2$ as easily verified in
 Ref.~\cite{illa2013transient}. Therefore, extrapolating the minimum value for $\langle \dot{\gamma} \rangle$ using this $k$ suggests that
 shear localization should occur trivially due to the start-up inertia for any real complex fluids in the experimentally accessible shear rates (typically $\langle \dot{\gamma} \rangle_{min} \sim 10^{-2}$ (1/s)).}
 \label{fig:fig6}
\end{figure}

Finally, Fig.~\ref{fig:fig6} provides a semilogarithmic "phase" diagram, in which the regions of localized and uniform shear are displayed.
The boundary indicates the minimum triggering value of $\langle \dot{\gamma} \rangle$,
at each $k$ required to yield significant heterogeneous structure relaxation, triggered by the start-up stress overshoot due to
the fluid inertia.
Above this value, a TSB forms, and below it the coupling between the structure relaxation and the local shear heterogeneity
during the start-up inertia is not strong enough to trigger the TSB. At this regime the
 structure relaxes homogeneously, manifested by uniform shear. The boundary between these two regimes moves to lower shear rates, as the
 relaxation exponent is increased.
As Ref.~\cite{illa2013transient} explains, in the framework of this model, a kinetic exponent on the order of $k \sim 2$ would be required 
to match the duration of the TSB to experiments. Unfortunately, using here an exponent as high as this would not be permitted
due to numerical and computational limitations (the number of required iterations for convergence).
Extrapolating the monotonically decreasing trend based on the computed values of Fig.~\ref{fig:fig5} to $k = 2$ hints that 
the minimum value for $\langle \dot{\gamma} \rangle$ should be well below the typical (minimum) experimentally used
shear rates ($\langle \dot{\gamma} \rangle_{min} \sim 10^{-2}$). Therefore, we are inclined to argue based on the model,
TSB formation and the subsequent localized flow dynamics due to start-up inertia only should be recovered trivially for all
real-world complex fluids. This has the important implication for the experiments that TSB should appear in all simple yield stress fluids with such
fluidization exponents irrespective of the measuring geometry. 

\section{Conclusions}
Earlier theoretical work proposes a scenario, where the transient shear banding is initialized by gradients in
stress~\cite{illa2013transient}, or inhomogeneity of the sample~\cite{adams2011transient, hinkle2015rate,moorcroft2013criteria}. The present study brings a new mechanism, which shows
that transient shear banding can result even due to the start-up stress gradients caused by accelerating flow. As we show here, 
depending on the linearity of the fluids' viscosity response, and its initial state, this effect can be either negligible or very significant.
When the initial viscosity of the fluid is small, and/or the time-dependence of the fluid is close to linear 
the stress gradient is not able to induce any inhomogeneities during the start-up flow.
However, according to our simulations,  when the fluid at rest is in a high viscosity state 
and shows time-dependence under constant shear rate with a large fluidization exponent
the start-up stress inhomogeneities initialize long lasting transient shear bands at all
experimentally accessible shear rates.
This occurs despite the fact that the influence of the Navier-Stokes instability is multiple orders of magnitude faster compared to the actual TSB. However, it creates a small but non-negligible inhomogeneity in the structure, which is amplified over time into a complete TSB.
An example of such material is the carbopol, a typical example of so called simple yield stress fluids, for which
the fluidization exponent is larger than 2~\cite{divoux2010transient} and the viscosity diverges at rest.
Our findings give a reasonable explanation for the initialization of transient shear banding even in experimental geometries
exhibiting negligible stress heterogeneities, such as the cone and plate one.
This suggests that stress heterogeneities should be anticipated in all
strongly shear thinning fluids during the start-up phase of rheological experiments.

\section{Acknowledgements}
This work was supported by the Academy of Finland through the COMP center of
excellence and the project number 278367. The simulations were performed using the
computer resources within the Aalto University School of Science “Science-IT” project.

%

\end{document}